%% file: main.tex
\documentclass[11pt]{article}

\usepackage[main, preprint]{neurips_2026}
\input{packages}

\title{GigaSpeechBench: A Real-World Multilingual Speech-to-Text Benchmark}

\author{
\textbf{Yujie Tu\textsuperscript{2,8,9}},
\textbf{Yifan Yang\textsuperscript{1}},
\textbf{Tianrui Wang\textsuperscript{4}},
\textbf{Yanqiao Zhu\textsuperscript{1}},
\textbf{Guodong Lin\textsuperscript{5}},
\textbf{Mingchen Shao\textsuperscript{6}}\\
\textbf{Haoran Wang\textsuperscript{1}},
\textbf{Junzhe Liu\textsuperscript{1}},
\textbf{Yuxiang Fu\textsuperscript{5}},
\textbf{Yizhou Peng\textsuperscript{7}},
\textbf{Changsong Liu\textsuperscript{7}},
\textbf{Peng Wang\textsuperscript{11}}\\
\textbf{Zhikang Niu\textsuperscript{1}},
\textbf{Yunchong Xiao\textsuperscript{3}},
\textbf{Haolong Zheng\textsuperscript{10}},
\textbf{Xiuwen Zheng\textsuperscript{10}},
\textbf{Xulin Fan\textsuperscript{10}}\\
\textbf{Wei-Qiang Zhang\textsuperscript{5,16}},
\textbf{Lei Xie\textsuperscript{6,15}},
\textbf{Longbiao Wang\textsuperscript{4}},
\textbf{Eng-Siong Chng\textsuperscript{7}},
\textbf{Jiajun Zhang\textsuperscript{8,9}}\\
\textbf{Kele Xu\textsuperscript{13}},
\textbf{Jianwei Yu\textsuperscript{3}},
\textbf{Binbin Zhang\textsuperscript{3,15}},
\textbf{Jiayu Du\textsuperscript{16}},
\textbf{Wupeng Wang\textsuperscript{3}},
\textbf{Zhigao Chen\textsuperscript{3}}\\
\textbf{Yuzhong Wu\textsuperscript{3}},
\textbf{Zhendong Peng\textsuperscript{3}},
\textbf{Bin Ma\textsuperscript{3}},
\textbf{Guoguo Chen\textsuperscript{14,16}},
\textbf{Xipeng Qiu\textsuperscript{2,12}},\\
\textbf{Mark Hasegawa-Johnson\textsuperscript{10}},
\textbf{Kai Yu\textsuperscript{1}},
\textbf{Zhifu Gao\textsuperscript{3}},
\textbf{Xiangang Li\textsuperscript{3}},
\textbf{Xie Chen\textsuperscript{1,2,16$\ddagger$}}
\\
\textsuperscript{1}SJTU
\textsuperscript{2}SII
\textsuperscript{3}Alibaba
\textsuperscript{4}TJU
\textsuperscript{5}THU
\textsuperscript{6}ASLP@NPU
\textsuperscript{7}NTU
\textsuperscript{8}CASIA
\textsuperscript{9}UCAS\\
\textsuperscript{10}UIUC
\textsuperscript{11}CUHK-SZ
\textsuperscript{12}FDU
\textsuperscript{13}CCSE
\textsuperscript{14}Seasalt.ai
\textsuperscript{15}WeNet
\textsuperscript{16}SpeechColab
\\[0.6em]
\small
\faGithub\ 
\href{https://github.com/SpeechColab/GigaSpeechBench}{\textnormal{https://github.com/SpeechColab/GigaSpeechBench}}
\\[0.2em]
\raisebox{-0.05em}{\includegraphics[height=0.75em]{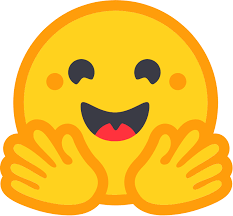}}\ 
\href{https://huggingface.co/datasets/speechcolab/GigaSpeechBench}{\textnormal{https://huggingface.co/datasets/speechcolab/GigaSpeechBench}}
}

\begin{document}
\maketitle
\input{text/abstract}
\input{text/introduction}

\input{text/relatedwork}
\input{text/Data}
\input{text/bench}
\input{text/conclusion}

\clearpage
\input{text/limitations}
\input{text/ethicsstatement}

\bibliographystyle{acl_natbib}
\bibliography{custom}

\appendix
\input{text/appendix}

\end{document}

%% file: packages.tex
\usepackage[utf8]{inputenc} 
\usepackage[T1]{fontenc}    
\usepackage{booktabs}       
\usepackage{amsfonts}       
\usepackage{nicefrac}       
\usepackage{microtype}      
\usepackage{xcolor}         

\usepackage{fontawesome5}

\usepackage{color,xcolor}
\usepackage{epsfig}
\usepackage{graphicx}
\usepackage{lipsum}
\newcommand{\cmark}{\ding{51}}
\newcommand{\xmark}{\ding{55}}

\usepackage{array}
\usepackage{booktabs}
\usepackage{colortbl}
\usepackage{multirow}
\usepackage{float}
\usepackage{caption}
\usepackage{adjustbox}
\usepackage{subcaption}
\usepackage{footnote}
\makesavenoteenv{tabular}
\makesavenoteenv{table}
\usepackage{makecell}
\usepackage{tablefootnote}
\usepackage{threeparttable}

\usepackage{amsmath,amsfonts,amssymb,bm}
\usepackage[super]{nth}

\usepackage{changepage}
\usepackage{extramarks}
\usepackage{lastpage}
\usepackage{setspace}
\usepackage{soul}
\usepackage{xspace}
\usepackage{wrapfig}

\usepackage{url}
\usepackage{hyperref}
\hypersetup{colorlinks=True, urlcolor=black}

\usepackage[ruled,vlined]{algorithm2e}
\usepackage{enumitem}
\usepackage{verbatim}
\usepackage{pifont}
\usepackage[acronyms]{glossaries}
\glsdisablehyper

%% file: text/abstract.tex
\begin{abstract}
While modern ASR systems achieve low error rates on high-resource benchmarks, such performance often overestimates real-world robustness. Existing evaluations address challenges in isolation, lacking a unified benchmark for domain terminology, age variation, dialects, accents, and low-resource languages, particularly across the Middle East and Southeast Asia, representing over one billion under-evaluated speakers. To address this gap, we introduce \textbf{GigaSpeechBench}, a comprehensive multilingual and multidimensional in-the-wild ASR \& AST benchmark comprising 690 hours of human-annotated speech. It features five modules: (1)~12 low-resource Middle Eastern and Southeast Asian languages, plus challenging Japanese and Korean; (2)~6 Chinese dialects; (3)~6 English accents; (4)~dense terminology across 12 vertical domains for Chinese and English; and (5)~older adult and child speech. We further provide human-annotated Chinese and English translations for 11 languages to support AST evaluation. Extensive evaluations of leading foundation models and commercial APIs reveal significant performance degradation in these challenging settings, exposing critical evaluation blind spots.
\end{abstract}

%% file: text/introduction.tex
\begin{figure}[t]
\centering
\includegraphics[width=0.8\linewidth]{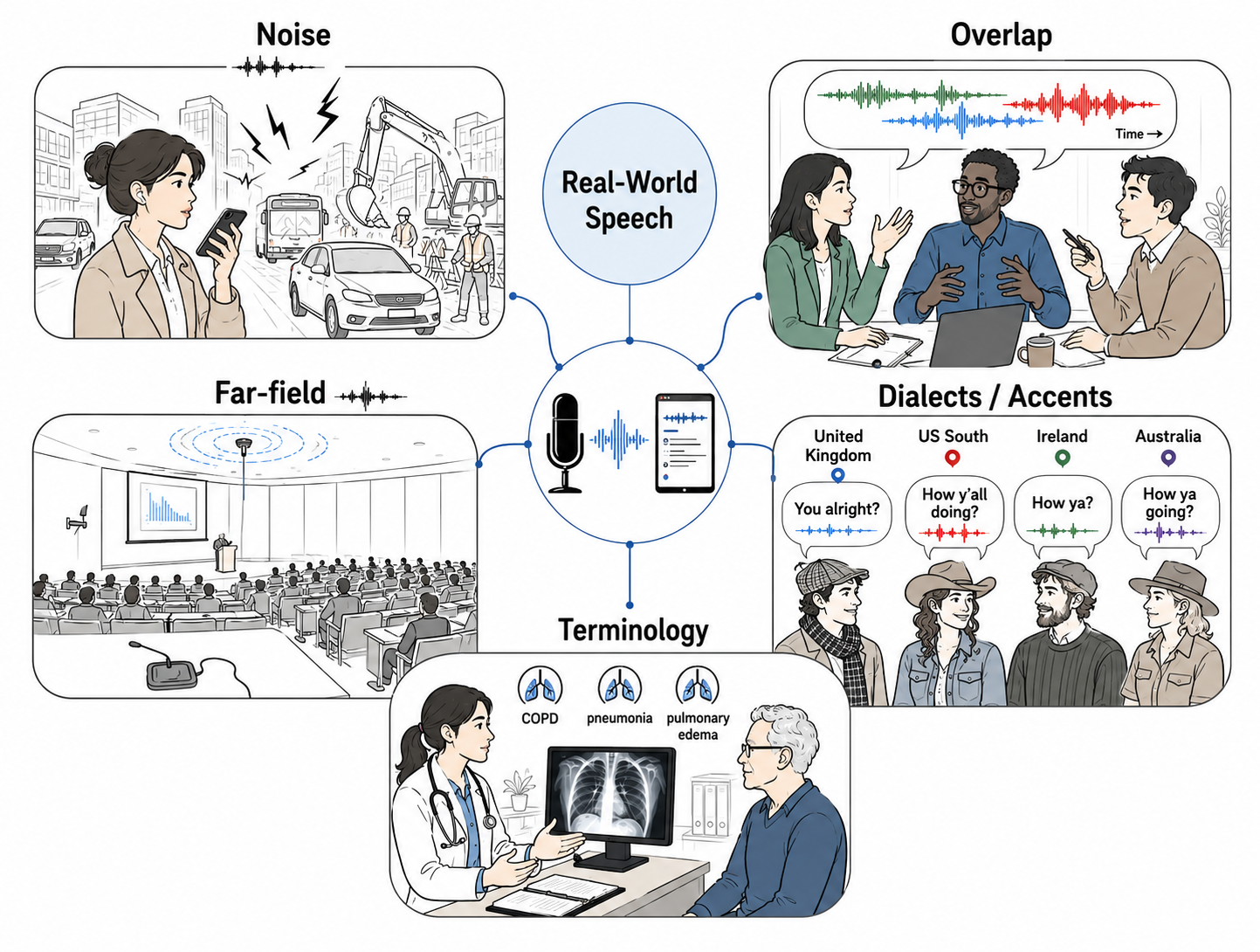}
\vspace{-1em}
\caption{
Real-world speech presents diverse acoustic, linguistic, and lexical challenges, ranging from noise, speaker overlap, and far-field recordings to dialects, accents, and domain-specific terminology.
}
\vspace{-1em}
\label{fig:teaser}
\end{figure}

\section{Introduction}
Automatic speech recognition (ASR) has advanced rapidly in recent years, driven by large-scale speech foundation models trained on massive multilingual corpora~\cite{whisper,gpt4o,Gemini25}.
Both open-source and commercial systems have achieved strong performance on widely used benchmarks, especially for high-resource languages such as English and Mandarin in controlled evaluation settings.
As widely used ASR benchmarks~\cite{commonvoice,fleurs} increasingly enter low-WER regimes, further gains on these datasets become less informative about real-world robustness.
Such low WERs and marginal gains increasingly reflect benchmark saturation, rather than reliable recognition under realistic acoustic and linguistic variability.

Existing multilingual ASR evaluations~\cite{commonvoice,fleurs,mls} typically prioritize language coverage over acoustic diversity, with test sets dominated by read or prompted speech collected under relatively clean conditions.
Such settings underrepresent key sources of acoustic variability, including spontaneous conversations, overlapping speech, background noise, far-field speech, and device-dependent recordings, as illustrated in Figure~\ref{fig:teaser}.

Beyond acoustic variability, linguistic and lexical variation also remain insufficiently evaluated.
Standard benchmarks provide limited coverage of regional dialects, non-standard varieties, and accented pronunciations~\cite{tang2021kespeech, sanabria2023edinburgh}.
Evaluation resources are also unevenly distributed, with languages across the Middle East and Southeast Asia comparatively under-evaluated despite their large speaker populations~\cite{wang2025openarabicasr, gigaspeech2}. 
In addition, current ASR evaluations remain limited in their coverage of terminology-dense speech, despite its importance in professional domains such as medicine, law, finance, and technology~\cite{wang2025contextasr}. 
Older adult and child speech requires dedicated ASR evaluation as age-related speech variation introduces unique challenges, including vocal tract development, higher pitch, and inconsistent pronunciation in children, as well as reduced volume, slower articulation, and tremors in older adults~\cite{zhou2025childmandarin,wang2025wildelder}.


Collectively, these limitations create a gap between benchmark performance and real-world ASR reliability, making it unclear whether recent systems are genuinely robust or merely well-adapted to existing benchmark distributions.
To address this gap, we introduce \textbf{GigaSpeechBench}, a 690-hour human-annotated ASR benchmark built from in-the-wild speech under complex acoustic conditions, organized into five modules:
\begin{itemize}[itemsep=0pt, topsep=2pt, leftmargin=*]
    \item \textbf{14 languages and regions:} 7 Arabic-speaking regions, including Iraq, Algeria, the United Arab Emirates, Egypt, Morocco, Saudi Arabia, and Syria; 5 Southeast Asian languages, including Indonesian, Malay, Filipino (Tagalog), Vietnamese, and Thai; and 2 East Asian languages with challenging speech, Japanese and Korean, with 20 hours each. For 11 of these languages, Chinese and English translation references are also provided for speech-to-text translation evaluation.
    \item \textbf{6 Chinese dialects:} Xiang, Jin, Gan, Min, Yue, and Wu, with 10 hours each.
    \item \textbf{6 English accents:} Chinese, Indian, Japanese, Filipino, Scottish, and Singaporean English, with 10 hours each.
    \item \textbf{12 terminology domains:} Agriculture, AI, Arts, Biotechnology, E-commerce, Engineering, Entertainment, Finance, Humanities, Law, Medicine, and Military, each with 10 hours of Chinese and 10 hours of English.
    \item \textbf{2 age groups:} Older adult and child speech, each with 10 hours of Chinese and 10 hours of English.
\end{itemize}

Evaluating leading foundation models and commercial APIs on GigaSpeechBench reveals that strong performance on existing benchmarks does not reliably transfer to these challenging settings, exposing critical evaluation blind spots. All resources will be released to facilitate reproducible, real-world ASR evaluation.

%% file: text/relatedwork.tex
\input{table/bench-compare}

\section{Related Work}
\subsection{Multilingual and Low Resource Benchmarks}
Multilingual evaluations such as FLEURS~\cite{fleurs} and Common Voice~\cite{commonvoice} typically rely on read speech or on standardized varieties with unstable splits. Other benchmarks address real-world needs but remain limited in scope: BABEL~\cite{gales2014speech} is narrowband; MLS~\cite{mls} and Open ASR Leaderboard~\cite{srivastav2025openasrleaderboard} (utilizing CoVoST-2~\cite{wang21s_interspeech}, etc.) cover a few European languages; ML-SUPERB~\cite{shi2023findings} focuses on representation learning; MLC-SLM~\cite{mu2026mlcslm} provides limited data per variety in clean conditions; and Open Universal Arabic ASR~\cite{wang2025openarabicasr} (pooling SADA~\cite{alharbi2024sada}, MASC~\cite{al2023masc}, MGB-2~\cite{ali2016mgb}) obscures per-dialect performance. In contrast, GigaSpeechBench provides 20 hours of human-annotated, real-world speech per language, explicitly targeting underrepresented Middle Eastern and Southeast Asian varieties.

\subsection{Chinese Dialects and Accented English}


Chinese dialect resources such as KeSpeech~\cite{tang2021kespeech}, the WenetSpeech series~\cite{wenetspeech-yue, wenetspeech-chuan, wenetspeech-wu}, MinSpeech~\cite{minspeech}, and YuBao ~\cite{chang2026towards} are either fragmented, focused on identification/retrieval, or lack a unified ASR evaluation protocol. For accented English, benchmarks such as EdAcc~\cite{sanabria2023edinburgh} reveal severe performance drops across diverse accents. GigaSpeechBench enables direct comparison by evaluating six Chinese dialects and six English accents under a unified 10-hour-per-variety protocol.

\subsection{Domain-Specific Terminology}
Average WER on general conversational data tends to obscure model struggles with dense, domain-specific vocabulary. ProfASR-Bench~\cite{piskala2025profasr} explicitly evaluates context-conditioned ASR across finance, medicine, legal, and technology domains, and identifies a context-utilization gap (CUG): even when paired with profile, domain, or oracle prompts, current Whisper-style~\cite{whisper} and audio-LM-based systems show little to no average-WER change and only modest, model-dependent gains on entity-rich tokens.
In contrast, GigaSpeechBench provides paired hotword lists across twelve vertical domains for both Chinese and English on genuine human speech, enabling controlled measurement of entity-aware ASR rather than relying on synthetic audio.

\subsection{Speaker Demographics}
Mainstream evaluations typically marginalize age-related acoustic extremes. Existing corpora address demographics in isolation, focusing either exclusively on children, such as MyST~\cite{pradhan2024my}, OGI Kids~\cite{shobaki2000ogi}, and ChildMandarin~\cite{zhou2025childmandarin}, or on seniors, such as SeniorTalk~\cite{NEURIPS2025_9131c4dc}.
GigaSpeechBench unifies both, offering 10 hours of child and older-adult speech in both Mandarin and English.

\subsection{Speech-to-Text Translation}
Public ST corpora (MuST-C~\cite{di2019must}, Europarl-ST~\cite{iranzo2020europarl}, CoVoST~2~\cite{wang21s_interspeech}, FLEURS~\cite{fleurs}, CS-FLEURS~\cite{yan2025cs}) are largely confined to European languages, formal settings, or synthetic/read audio. GigaSpeechBench advances ST evaluation by providing human-translated references on the \emph{same} in-the-wild audio used for ASR, covering 11 underrepresented Middle Eastern and Southeast Asian languages under noisy, multi-speaker conditions.

%% file: table/bench-compare.tex
\begin{table*}[!t]
\centering
\caption{Comparison of open-source ASR benchmarks across multilingual, Chinese dialectal, accented English, terminology, and age-variation evaluation settings. \#Lang./Var./Dom./Age denotes the number of languages, varieties, domains, or age groups. Hours per Lang./Var./Dom./Age denotes test-set hours per language, variety, domain, or age group. \textsuperscript{*}denotes benchmarks reusing existing open-source speech corpora.}
\renewcommand{\arraystretch}{1.1}
\setlength{\tabcolsep}{3pt}
\resizebox{\linewidth}{!}{
\begin{tabular}{lcccccccc}
\toprule[1pt]
Benchmark
& \makecell[c]{Multilingual}
& Dialect
& Accent
& Terminology
& \makecell[c]{Age\\Variation}
& \makecell[c]{\#Lang./Var.\\Dom./Age}
& \makecell[c]{Hours per\\Lang./Var.\\Dom./Age}
& Speech Type \\
\midrule

FLEURS~\cite{fleurs}
& \cmark & \xmark & \xmark & \xmark & \xmark
& 102
& <3
& Read \\

Common Voice~\cite{commonvoice}
& \cmark & \xmark & \xmark & \xmark & \xmark
& 29
& -
& Read \\

MLS~\cite{mls}
& \cmark & \xmark & \xmark & \xmark & \xmark
& 8
& 9.3
& Read \\

BABEL~\cite{gales2014speech}
& \cmark & \xmark & \xmark & \xmark & \xmark
& 25
& -
& Spontaneous \\

MLC-SLM~\cite{mu2026mlcslm}
& \cmark & \xmark & \cmark & \xmark & \xmark
& 11
& $\sim$2
& Spontaneous \\

ML-SUPERB~\cite{mlsuperb}\textsuperscript{*}
& \cmark & \xmark & \xmark & \xmark & \xmark
& 143
& <1
& Mixed \\

Open ASR Leaderboard~\cite{srivastav2025openasrleaderboard}\textsuperscript{*}
& \cmark & \xmark & \xmark & \xmark & \xmark
& 5
& -
& Mixed \\

Open Universal Arabic ASR~\cite{wang2025openarabicasr}\textsuperscript{*}
& \cmark & \xmark & \xmark & \xmark & \xmark
& -
& -
& Mixed \\

KeSpeech~\cite{tang2021kespeech}
& \xmark & \cmark & \xmark & \xmark & \xmark
& 8
& -
& Read \\

ContextASR-Bench~\cite{wang2025contextasr}
& \xmark & \xmark & \xmark & \cmark & \xmark
& 10+
& -
& Synthetic \\

\midrule

\textbf{GigaSpeechBench} (Ours)
& \cmark & \cmark & \cmark & \cmark & \cmark
& \makecell[c]{14/6/6/12/2}
& \makecell[c]{20/10/10/20/10}
& Spontaneous \\

\bottomrule[1pt]
\end{tabular}
}
\label{tab:benchmark_positioning}
\end{table*}

%% file: text/Data.tex
\input{table/low-resource-language}
\input{table/Common-Voice}
\input{table/fleurs}
\input{table/English-Dialects}
\input{table/Chinese-Dialects}
\input{table/Chinese_BCER}
\input{table/English_BWER}
\input{table/Old-Child}
\input{table/Translate-EN}
\input{table/Translate-CH}

\section{GigaSpeechBench Construction}

We construct GigaSpeechBench through the following curation pipeline designed to collect spontaneous speech in the target language or regional variety, cover diverse acoustic and linguistic conditions, and ensure reliable manual transcription for modern ASR evaluation.

\paragraph{Source Discovery and Video Selection.}
We first build candidate source pools from YouTube. Since the language or regional variety of a video cannot be determined from metadata alone, we use a heuristic screening procedure based on multiple evidence sources, including channel descriptions, video titles, comments, uploader information, and other publicly available metadata. A channel is retained only when these signals consistently suggest that it contains speech from the target language or regional variety.
From the selected channels, we prioritize videos containing \textbf{spontaneous conversational speech} and exclude recordings dominated by read, scripted, or narration-style speech. When sufficient data is available, we prefer \textbf{recently published videos} to reduce potential overlap with existing model pre-training data.

We then perform video-level audio screening before annotation. Videos longer than one hour are removed to prevent a few speakers or sources from dominating the benchmark and to limit within-source acoustic heterogeneity. We also exclude recordings in which speech is largely unintelligible due to severe noise, distortion, background music, or persistent speaker overlap.
We do not remove all acoustically challenging samples. Instead, we retain recordings with natural background noise, far-field speech, channel variation, and occasional speaker overlap, as long as the target speech remains dominant and transcribable. This criterion allows the benchmark to reflect real-world ASR conditions while avoiding samples that cannot be reliably annotated or evaluated.

\paragraph{Segmentation and Manual Transcription.}
The screened videos are sent to a professional annotation company for voice activity detection and manual transcription. Annotators first segment the continuous audio into utterance-level speech segments. Segment boundaries are placed near low-energy points in the waveform.

Each retained segment is transcribed in the native writing form of the target language or regional variety. Annotators are instructed to transcribe only target-language speech. Segments are marked as invalid if they contain no speech, pure background music, unintelligible speech, speech fully masked by noise, or speech primarily in a non-target language. For mixed-language or overlapping-speech cases, a segment is retained only when the target-language content is clear enough for reliable native ASR evaluation.

\paragraph{Quality Control and Test Set Construction.}
After annotation, the annotation company performs manual quality inspection on the transcriptions. The retained annotations achieve a reported transcription accuracy above 98\% according to the provider's quality report.
We further conduct post-processing before forming the final benchmark. This step removes invalid segments missed during annotation, segments dominated by non-target-language speech, incomplete or clearly mismatched transcriptions, and audio whose intelligibility is too low for stable evaluation. We also exclude segments shorter than 0.5 seconds from metric computation, since such utterances often contain too few lexical units and can lead to unstable ASR error estimates.

The final benchmark consists of utterance-level audio segments paired with native manual transcripts. These test sets are used for native ASR evaluation under diverse but transcribable acoustic conditions. We report the segment duration distribution and word-count distribution in Appendix~\ref{sec:appendix_duration}.

%% file: table/low-resource-language.tex
\begin{table*}[!t]
\centering
\caption{Low-resource languages benchmark results. WER (\%) is reported for Arabic and Southeast Asian languages, while CER (\%) is reported for East Asian languages.}
\renewcommand\tabcolsep{3pt}
\renewcommand{\arraystretch}{1.1}
\resizebox{\textwidth}{!}{
\begin{tabular}{l*{14}{c}}
\toprule[1pt]

\multirow{2}{*}{\textbf{System}}
& \multicolumn{2}{c}{\textbf{East Asia}} 
& \multicolumn{5}{c}{\textbf{Southeast Asia}} 
& \multicolumn{7}{c}{\textbf{Arabic}} \\

\cmidrule(lr){2-3} \cmidrule(lr){4-8} \cmidrule(l){9-15}

& JPN & KOR
& IDN & MYS & PHL & VNM & THA
& IRQ & DZA & ARE & EGY & MAR & SAU & SYR \\

\midrule

Azure
& 27.51 & 13.13
& 25.50 & 35.20 & \underline{26.08} & 10.95 & 15.66
& 34.61 & 51.22 & 42.82 & 47.65 & 56.64 & 20.09 & 17.74 \\

Chirp 3
& 36.22 & 15.96
& 19.98 & 29.04 & 28.18 & \textbf{9.63} & 17.52
& 35.71 & 53.11 & 42.88 & 42.71 & 52.30 & \underline{16.76} & 24.13 \\

ElevenLabs Scribe v2
& 29.95 & \underline{11.81}
& 22.91 & 38.52 & 27.15 & 10.52 & \underline{13.90}
& 38.67 & 50.43 & 46.10 & 44.44 & 60.06 & 33.33 & 14.73 \\

Meta OmniASR 3B
& 58.74 & 26.76
& 37.91 & 68.79 & 45.03 & 19.60 & 30.72
& 38.80 & 57.68 & 50.83 & 52.37 & 65.52 & 25.31 & 17.86 \\

Qwen3-ASR-Flash
& 28.40 & 17.52
& 20.45 & 60.18 & 47.83 & 11.31 & 17.08
& \underline{33.21} & 57.18 & 44.24 & 48.78 & 68.51 & 19.21 & 14.41 \\

Qwen3-ASR-1.7B
& 31.77 & 12.90
& 22.29 & 50.68 & 51.58 & 11.90 & 15.14
& 41.27 & 63.43 & 53.22 & 59.23 & 76.65 & 25.85 & 18.50 \\

NVIDIA NeMo
& 32.31 & --
& -- & -- & -- & -- & --
& 43.22 & 62.66 & 56.00 & 54.83 & 73.65 & 29.28 & 20.13 \\

GPT-4o Transcribe
& 44.34 & 41.31
& 37.95 & 52.30 & 38.60 & 29.24 & 48.78
& 54.53 & 63.14 & \textbf{26.26} & 64.23 & 71.26 & 42.38 & 31.67 \\

Gemini 3.0 Flash
& 39.84 & 16.78
& 24.18 & 40.92 & 29.17 & 11.69 & 26.58
& 36.55 & \textbf{44.22} & 45.06 & \underline{41.22} & \underline{51.99} & 20.10 & \underline{14.40} \\

Whisper Large v3
& 39.28 & 18.53
& 27.40 & 46.15 & 30.88 & 18.17 & 27.02
& 51.04 & 72.02 & 68.41 & 69.78 & 91.89 & 32.79 & 19.12 \\

Dolphin Small
& 40.30 & 39.05
& 32.53 & 52.19 & 61.08 & 21.68 & 24.40
& 62.05 & 72.44 & 75.62 & 74.70 & 75.96 & 50.91 & 30.03 \\

Dolphin Base
& 39.61 & 28.59
& 31.29 & 54.24 & 68.36 & 21.59 & 26.97
& 65.20 & 78.26 & 82.87 & 85.31 & 89.74 & 52.35 & 38.12 \\

FunASR-MLT-Nano
& 29.03 & 16.57
& 27.68 & 43.01 & 36.45 & 14.02 & 20.75
& -- & -- & -- & -- & -- & -- & -- \\

FunASR-Realtime
& \textbf{25.44} & \textbf{9.92}
& \textbf{14.87} & \textbf{25.20} & \textbf{23.69} & \underline{9.75} & \textbf{10.76}
& 53.44 & 66.30 & 66.70 & 63.33 & 74.10 & 37.67 & 24.24 \\

Qwen3.5-Omni-Plus
& \underline{27.36} & 13.10
& \underline{18.05} & \underline{28.78} & 26.21 & 9.90 & 15.10
& \textbf{28.54} & \underline{47.11} & \underline{35.15} & \textbf{37.12} & \textbf{51.34} & \textbf{16.56} & \textbf{13.76} \\

Deepgram Nova 3
& -- & --
& -- & -- & -- & -- & --
& 47.54 & 57.90 & 52.06 & 52.77 & 60.00 & 25.02 & 30.61 \\

\midrule

\textbf{Best}
& \textbf{25.44} & \textbf{9.92}
& \textbf{14.87} & \textbf{25.20} & \textbf{23.69} & \textbf{9.63} & \textbf{10.76}
& \textbf{28.54} & \textbf{44.22} & \textbf{26.26} & \textbf{37.12} & \textbf{51.34} & \textbf{16.56} & \textbf{13.76} \\

\bottomrule[1pt]
\end{tabular}
}
\label{tab:low_resource_main}
\end{table*}

%% file: table/Common-Voice.tex
\begin{table}[t]
\centering
\caption{Common Voice benchmark results. Only languages with available evaluation data are reported. WER (\%) is used for alphabetic languages, and CER (\%) for logographic languages.}
\begin{tabular}{lcccccc}
\toprule[1pt]

\textbf{System}
& AR & IDN & VNM & THA & JPN & KOR \\

\midrule

Azure
& 13.09 & 10.69 & 10.1 & 4.45 & 21.71 & 7.14 \\

Chirp3
& -- & 5.45 & \textbf{7.00} & 5.52 & 24.86 & 5.55 \\

Elevenlabs\_scribe\_v2
& 10.14 & \textbf{3.87} & 8.55 & \textbf{1.50} & 18.84 & 4.48 \\

OmniASR\_LLM\_3B 
& 8.96 & 10.36 & 16.08 & 5.04 & 25.53 & 12.16 \\

Qwen3-asr-flash
& 10.51 & 4.13 & 8.3 & 3.69 & 19.74 & 4.64 \\

Qwen3-asr-1.7B
& 17.05 & 6.92 & 12.31 & 3.73 & 23.09 & 7.22 \\

Nvidia-nemo
& \textbf{7.19} & -- & -- & -- & 19.74 & 23.83 \\

GPT-4o Transcribe
& 12.23 & 5.96 & 13.40 & 5.00 & 20.8 & 6.13 \\

Gemini 3.0 flash
& 10.81 & 5.74 & 10.67 & 3.83 & 25.37 & 9.42 \\

Whisper
& 15.47 & 8.13 & 14.71 & 6.84 & 22.54 & 5.84 \\

Dolphin\_small
& 21.77 & 8.55 & 14.39 & 4.78 & 21.76 & 7.72 \\

Dolphin\_base
& 35.79 & 12.52 & 22.23 & 6.46 & 24.73 & 11.08 \\

\midrule
\textbf{Best}
& \textbf{7.19} & \textbf{3.87} & \textbf{7.00} & \textbf{1.50} & \textbf{18.84} & \textbf{4.48} \\

\bottomrule[1pt]
\end{tabular}
\label{tab:commonvoice_final}
\end{table}

%% file: table/fleurs.tex
\begin{table}[!t]
\centering
\caption{FLEURS benchmark results. WER (\%) is reported for multilingual speech recognition, while CER (\%) is reported for East Asian languages.}
\begin{tabular}{lcccccccc}
\toprule[1pt]

\textbf{System}
& EGY & IDN & MYS & PHL & VNM & THA & JPN & KOR \\

\midrule

Azure
& 19.54 & 10.38 & 10.52 & 14.68 & 8.59 & 8.66 & 6.29 & 6.08 \\

Chirp3
& 14.16 & 3.39 & \textbf{3.88} & \textbf{6.61} & 3.19 & 10.38 & 3.76 & 4.31 \\

elevenlabs\_scribe\_v2
& 13.5 & \textbf{2.94} & 3.92 & 7.48 & \textbf{2.71} & 6.29 & \textbf{2.41} & \textbf{3.78} \\

OmniASR\_LLM\_3B
& \textbf{7.78} & 10.2 & 11.47 & 12.05 & 12.2 & 8.56 & 11.02 & 9.02 \\

Qwen3-asr-flash
& 15.88 & 4.96 & 12.35 & 19.28 & 3.73 & 6.55 & 3.71 & 4.29 \\

Qwen3-asr-1.7B
& 17.84 & 5.89 & 10.69 & 23.06 & 5.52 & 6.76 & 5.51 & 4.53 \\

Nvidia-nemo
& 16.8 & -- & -- & -- & -- & -- & 5.83 & 11.67 \\

GPT-4o Transcribe
& 14.28 & 4.13 & 4.43 & 7.27 & 3.88 & \textbf{5.88} & 3.48 & 4.19 \\

Gemini 3.0 flash
& 12.78 & 3.71 & 4.77 & 8.04 & 2.98 & 6.48 & 3.55 & 4.22 \\

Whisper
& 10.96 & 6.24 & 7.73 & 10.96 & 7.93 & 8.62 & 5.33 & 4.73 \\

Dolphin\_small
& 19.82 & 13.66 & 13.48 & 17.95 & 11.58 & 11.22 & 8.58 & 9.5 \\

Dolphin\_base
& 26.84 & 14.77 & 18 & 21.49 & 16.1 & 11.49 & 10.44 & 10.03 \\

\midrule

\textbf{Best}
& \textbf{7.78} & \textbf{2.94} & \textbf{3.88} & \textbf{6.61} & \textbf{2.71} & \textbf{5.88} & \textbf{2.41} & \textbf{3.78} \\

\bottomrule[1pt]
\end{tabular}
\label{tab:fleurs}
\end{table}

%% file: table/English-Dialects.tex
\begin{table}[!t]
\centering
\caption{English accent benchmark results, reported in WER (\%).}
\renewcommand\tabcolsep{2pt}
\begin{tabular}{lcccccc}
\toprule[1pt]

\textbf{System}
& CHN-EN
& IND-EN
& JPN-EN
& PHL-EN
& SCT-EN
& SGP-EN\\
\midrule

Azure
& 17.35 & 33.00 & 21.48 & 12.95 & 28.22 & 14.34 \\

Chirp 3
& 16.76 & 8.11 & 18.69 & 11.96 & 31.32 & 16.56 \\

ElevenLabs Scribe v2
& 18.69 & 11.76 & 25.60 & 20.13 & 36.44 & 18.75 \\

Meta OmniASR 3B
& 37.11 & 21.67 & 44.21 & 44.65 & 54.22 & 44.18 \\

Qwen3-ASR-Flash
& 16.67 & 11.66 & 20.23 & 14.22 & \textbf{23.68} & 12.49 \\

Qwen3-ASR-1.7B
& 14.62 & \textbf{7.04} & 21.52 & \textbf{10.81} & 24.29 & \textbf{12.44} \\

NVIDIA NeMo
& 24.38 & 10.56 & 29.25 & 21.29 & 41.64 & 17.45 \\

GPT-4o Transcribe
& 66.12 & 17.12 & 38.57 & 38.04 & 43.57 & 35.85 \\

Gemini 3.0 Flash
& 20.64 & 8.31 & 30.43 & 15.79 & 34.43 & 16.34 \\

Whisper Large v3
& 17.28 & 7.91 & 17.55 & 13.68 & 27.24 & 14.02 \\

FunASR-MLT-Nano
& 17.84 & 8.51 & 72.09 & 14.00 & 33.92 & 14.28 \\

FunASR-Realtime
& \underline{13.27} & 7.70 & \textbf{15.25} & \underline{11.10} & 26.60 & \underline{12.45} \\

Qwen3.5-Omni-Plus
& \textbf{12.98} & \underline{7.21} & \underline{15.67} & 11.73 & \underline{24.27} & 12.47 \\

BigASR
& 14.13 & 9.22 & 24.81 & 13.99 & 35.83 & 15.86 \\

SeedASR
& 14.09 & 9.21 & 24.79 & 14.00 & 35.82 & 15.86 \\

\midrule

\textbf{Best}
& \textbf{12.98} & \textbf{7.04} & \textbf{15.25} & \textbf{10.81} & \textbf{23.68} & \textbf{12.44} \\

\bottomrule[1pt]
\end{tabular}
\label{tab:english_accents}
\end{table}

%% file: table/Chinese-Dialects.tex
\begin{table}[!t]
\centering
\caption{Chinese dialect benchmark results, reported in CER (\%).}
\begin{tabular}{lcccccc}
\toprule[1pt]

\textbf{System}
& XIANG
& JIN
& GAN
& MIN
& YUE
& WU \\

\midrule

Azure
& 43.26 & 36.48 & 58.37 & 67.20 & 11.77 & 33.70 \\

Chirp 3
& 71.88 & 59.38 & 71.06 & 89.34 & 47.70 & 85.23 \\

ElevenLabs Scribe v2
& 54.27 & 44.86 & 68.39 & 71.41 & 32.09 & 65.45 \\

Meta OmniASR 3B
& 62.77 & 50.98 & 65.79 & 90.17 & 48.36 & 73.68 \\

Qwen3-ASR-Flash
& 27.38 & 31.68 & 47.32 & 59.60 & 11.63 & 31.93 \\

Qwen3-ASR-1.7B
& 25.01 & 27.62 & 49.48 & 56.98 & \underline{7.13} & \underline{24.20} \\

NVIDIA NeMo
& 85.49 & 80.16 & 83.69 & 94.74 & 95.44 & 86.42 \\

GPT-4o Transcribe
& 71.26 & 63.48 & 74.33 & 69.95 & 19.29 & 74.59 \\

Gemini 3.0 Flash
& 116.02 & 61.23 & 73.42 & 74.87 & 24.39 & 72.35 \\

Whisper Large v3
& 60.58 & 53.78 & 66.13 & 69.14 & 39.75 & 73.32 \\

Dolphin Small
& 37.08 & 32.67 & 60.60 & 59.45 & 23.86 & 25.77 \\

Dolphin Base
& 49.70 & 40.13 & 65.21 & 68.14 & 28.70 & 32.45 \\

FunASR-MLT-Nano
& 28.96 & 28.09 & 54.77 & 68.87 & 8.66 & 29.21 \\

FunASR-Realtime
& \textbf{19.92} & \textbf{22.83} & \textbf{43.20} & \textbf{27.72} & \textbf{6.13} & \textbf{16.96} \\

Qwen3.5-Omni-Plus
& \underline{21.52} & 24.19 & \underline{45.18} & 39.85 & 7.94 & 24.64 \\

BigASR
& 22.31 & \underline{23.81} & 53.63 & 36.85 & 10.54 & 31.28 \\

SeedASR
& 22.41 & 23.89 & 53.77 & \underline{33.99} & 10.30 & 32.11 \\

\midrule

\textbf{Best}
& \textbf{19.92} & \textbf{22.83} & \textbf{43.20} & \textbf{27.72} & \textbf{6.13} & \textbf{16.96} \\

\bottomrule[1pt]
\end{tabular}
\label{tab:chinese_dialects}
\end{table}

%% file: table/Chinese_BCER.tex
\begin{table*}[!t]
\centering
\caption{Chinese vertical-domain benchmark results, reported in B-CER (\%).}
\renewcommand\tabcolsep{2pt}
\resizebox{\linewidth}{!}{
\begin{tabular}{lcccccccccccc}
\toprule[1pt]

\textbf{Model}
& AGR-CH
& AIT-CH
& ART-CH
& BIO-CH
& ECM-CH
& ENG-CH
& ENT-CH
& FIN-CH
& HUM-CH
& LAW-CH
& MED-CH
& MIL-CH \\
\midrule

Duration (valid ref)
& 6.69h & 10.32h & 9.85h & 10.23h & 10.29h & 10.76h
& 8.14h & 10.64h & 10.28h & 10.16h & 10.12h & 10.33h \\

\midrule

Azure
& 35.98 & 32.83 & 27.17 & 30.12 & 37.71 & 37.76 & 34.52 & 9.97 & 37.97 & 14.21 & 26.04 & 7.44 \\

Chirp 3
& 33.06 & 32.92 & 25.96 & 30.05 & 42.10 & 25.39 & 47.54 & 11.30 & 31.00 & 42.92 & 24.02 & 11.50 \\

ElevenLabs Scribe v2
& 29.40 & 26.86 & 18.59 & 27.90 & 37.66 & 25.20 & 32.79 & 6.90 & 27.15 & 12.31 & 24.93 & 6.04 \\

Meta OmniASR 3B
& 54.40 & 58.14 & 35.54 & 45.07 & 46.85 & 48.02 & 52.78 & 19.24 & 41.88 & 19.11 & 47.62 & 21.15 \\

Qwen3-ASR-Flash
& 15.55 & 26.75 & 17.95 & 20.02 & 27.78 & 15.78 & 27.89 & 4.69 & 14.98 & 16.06 & 15.54 & 7.68 \\

Qwen3-ASR-1.7B
& 20.48 & 23.28 & 14.57 & 19.47 & 30.28 & 19.72 & 34.33 & 5.07 & 22.84 & 10.06 & 18.13 & 5.29 \\

NVIDIA NeMo
& 69.15 & 78.49 & 55.91 & 65.51 & 68.51 & 65.91 & 74.94 & 40.25 & 56.88 & 33.29 & 65.70 & 46.80 \\

GPT-4o Transcribe
& 38.49 & 32.46 & 30.67 & 33.27 & 53.67 & 48.55 & 46.97 & 17.90 & 29.30 & 18.87 & 30.93 & 11.06 \\

Gemini 3.0 Flash
& 21.50 & 18.43 & 17.44 & 20.42 & 38.12 & 18.25 & 39.90 & 7.26 & 18.40 & 14.71 & 15.62 & 6.79 \\

Whisper Large v3
& 47.07 & 36.10 & 33.15 & 37.45 & 45.13 & 39.39 & 50.03 & 16.95 & 39.91 & 19.67 & 42.22 & 13.81 \\

Dolphin Small
& 44.02 & 55.23 & 32.93 & 38.09 & 42.12 & 41.58 & 44.83 & 17.82 & 39.02 & 14.05 & 41.90 & 20.59 \\

Dolphin Base
& 51.83 & 62.50 & 38.04 & 45.86 & 46.77 & 46.53 & 53.57 & 25.05 & 43.06 & 17.48 & 47.62 & 26.71 \\

FunASR-MLT-Nano
& 27.91 & 31.69 & 18.38 & 26.47 & 30.86 & 25.85 & 31.25 & 6.81 & 31.44 & 10.97 & 22.92 & 6.37 \\

FunASR-Realtime
& \textbf{6.11} & \textbf{15.75} & \textbf{9.03} & \textbf{14.07} & \textbf{22.54} & \textbf{9.87} & \textbf{20.21} & \textbf{2.45} & \textbf{10.56} & \textbf{9.20} & \textbf{8.74} & \textbf{2.41} \\

Qwen3.5-Omni-Plus
& \underline{8.70} & \underline{18.09} & \underline{9.36} & \underline{14.10} & \underline{22.79} & \underline{12.51} & \underline{20.52} & \underline{2.95} & \underline{10.99} & \underline{9.85} & \underline{9.74} & \underline{3.08} \\

BigASR
& 12.88 & 20.26 & 12.01 & 20.50 & 23.09 & 17.26 & 22.46 & 4.61 & 15.89 & 10.72 & 14.42 & 5.05 \\

SeedASR
& 12.71 & 20.10 & 11.94 & 19.81 & 23.09 & 17.33 & 22.08 & 4.55 & 15.89 & 10.63 & 14.26 & 5.10 \\

\midrule

\textbf{MIN}
& \textbf{6.11} & \textbf{15.75} & \textbf{9.03} & \textbf{14.07} & \textbf{22.54} & \textbf{9.87} & \textbf{20.21} & \textbf{2.45} & \textbf{10.56} & \textbf{9.20} & \textbf{8.74} & \textbf{2.41} \\

\bottomrule[1pt]
\end{tabular}
}
\label{tab:vertical_domain_chinese_bcer}
\end{table*}

%% file: table/English_BWER.tex
\begin{table*}[!t]
\centering
\caption{English vertical-domain benchmark results, reported in B-WER (\%).}
\renewcommand\tabcolsep{4pt}
\resizebox{\linewidth}{!}{
\begin{tabular}{lcccccccccccc}
\toprule[1pt]

\textbf{Model}
& AGR-EN
& AIT-EN
& ART-EN
& BIO-EN
& ECM-EN
& ENG-EN
& ENT-EN
& FIN-EN
& HUM-EN
& LAW-EN
& MED-EN
& MIL-EN \\
\midrule

Duration (valid ref)
& 10.32h & 10.46h & 10.11h & 10.50h & 10.57h & 11.60h
& 10.12h & 11.43h & 10.03h & 10.32h & 10.31h & 11.20h \\

\midrule

Azure
& 9.84 & 28.56 & 12.02 & 18.52 & 19.45 & 14.74 & 21.19 & 13.17 & 7.79 & 17.78 & 14.88 & 19.76 \\

Chirp 3
& 7.92 & 27.63 & 8.58 & 15.12 & 16.13 & 13.28 & 21.16 & 12.05 & 6.25 & 15.97 & 13.10 & 19.06 \\

ElevenLabs Scribe v2
& 9.54 & 29.95 & 9.46 & 16.46 & 18.31 & 13.46 & 20.32 & 14.33 & 8.56 & 17.18 & 14.19 & 22.15 \\

Meta OmniASR 3B
& 22.69 & 42.15 & 22.60 & 28.45 & 27.60 & 26.53 & 40.29 & 22.74 & 12.22 & 27.87 & 24.59 & 21.03 \\

Qwen3-ASR-Flash
& \underline{6.91} & 26.54 & 14.61 & 15.85 & 14.54 & 11.32 & \underline{17.51} & \underline{10.27} & 8.48 & 19.93 & 14.65 & 18.15 \\

Qwen3-ASR-1.7B
& 7.36 & 26.79 & 8.20 & 15.67 & 15.71 & 11.70 & 19.42 & 10.35 & 6.12 & 15.13 & 13.52 & 15.79 \\

NVIDIA NeMo
& 14.50 & 31.73 & 17.18 & 24.07 & 23.98 & 18.50 & 36.01 & 13.34 & 10.19 & 20.66 & 14.36 & \textbf{11.53} \\

GPT-4o Transcribe
& 15.06 & 44.01 & 18.79 & 17.09 & 36.64 & 29.04 & 29.87 & 19.75 & 10.48 & 20.85 & 32.10 & 23.87 \\

Gemini 3.0 Flash
& 7.68 & 27.05 & \underline{7.04} & \underline{14.56} & 16.05 & 11.56 & 19.15 & 11.55 & 5.66 & \underline{13.03} & 12.45 & 18.24 \\

Whisper Large v3
& 7.60 & 28.53 & 9.00 & 16.34 & 16.26 & 12.64 & 19.47 & 11.64 & 6.11 & 15.64 & 13.78 & 18.99 \\

FunASR-MLT-Nano
& 10.02 & 28.24 & 10.96 & 19.33 & 17.45 & 12.93 & 25.42 & 12.78 & 7.79 & 18.75 & 15.30 & 18.14 \\

FunASR-Realtime
& 7.80 & \textbf{22.51} & 8.92 & 14.92 & \textbf{13.34} & \textbf{9.21} & 20.48 & \underline{10.27} & \textbf{5.05} & 14.95 & \underline{12.35} & \underline{15.47} \\

Qwen3.5-Omni-Plus
& \textbf{6.82} & \underline{25.83} & \textbf{6.47} & \textbf{14.33} & \underline{14.45} & \underline{10.23} & \textbf{16.93} & \textbf{10.24} & \underline{5.39} & \textbf{12.77} & \textbf{12.31} & 17.77 \\

BigASR
& 11.05 & 29.01 & 10.94 & 16.74 & 17.27 & 11.70 & 24.58 & 13.30 & 7.13 & 17.45 & 15.45 & 19.42 \\

SeedASR
& 12.79 & 29.48 & 11.13 & 16.83 & 17.17 & 11.62 & 24.97 & 13.20 & 7.19 & 17.48 & 15.36 & 19.46 \\

\midrule

\textbf{MIN}
& \textbf{6.82} & \textbf{22.51} & \textbf{6.47} & \textbf{14.33} & \textbf{13.34} & \textbf{9.21} & \textbf{16.93} & \textbf{10.24} & \textbf{5.05} & \textbf{12.77} & \textbf{12.31} & \textbf{11.53} \\

\bottomrule[1pt]
\end{tabular}
}
\label{tab:vertical_domain_english_bwer}
\end{table*}

%% file: table/Old-Child.tex
\begin{table}[t]
\centering
\caption{Child and elderly speech recognition results on the full four-subset benchmark using the revised reference transcripts. WER (\%) is used for English speech and CER (\%) for Chinese speech; lower is better. Bold and underlined scores denote the best and second-best results in each column, respectively, ranked using unrounded scores. A dash indicates that results are unavailable.}
\label{tab:child_old_asr_results}
\begin{tabular}{lcccc}
\toprule[1pt]
Model & CHILD-EN & CHILD-CH & OLD-EN & OLD-CH \\
\midrule
Valid reference duration & 9.98h & 9.83h & 10.09h & 10.04h \\
\midrule
Qwen Audio 3.1 ASR Flash & \textbf{17.91} & 9.86 & \underline{14.19} & \underline{9.75} \\
Qwen3-ASR-1.7B & \underline{18.26} & 10.34 & \textbf{13.54} & 10.43 \\
Qwen3.5-Omni-Plus & 19.89 & \textbf{9.30} & 15.01 & \textbf{9.13} \\
BigASR & 19.69 & \underline{9.42} & 15.17 & 9.77 \\
SeedASR & 19.67 & 9.42 & 15.19 & 9.78 \\
Fun-ASR-Nano-2512 & 22.95 & 13.47 & 16.70 & 13.75 \\
Azure & 23.10 & 15.58 & 15.71 & 12.85 \\
ElevenLabs Scribe v2 & 27.82 & 16.55 & 20.43 & 14.58 \\
Whisper Large v3 & 21.17 & 26.03 & 16.67 & 21.22 \\
Chirp 3 & 20.58 & 26.90 & 15.34 & 23.71 \\
Gemini 3.0 Flash & 24.08 & 31.18 & 19.74 & 26.01 \\
Meta OmniASR 3B & 45.21 & 23.95 & 24.22 & 20.46 \\
GPT-4o Transcribe & 35.86 & 33.59 & 36.85 & 36.02 \\
NVIDIA NeMo & 32.11 & 58.64 & 17.82 & 49.17 \\
Dolphin Small & - & 16.67 & - & 15.86 \\
Dolphin Base & - & 22.38 & - & 18.64 \\
\midrule
\textbf{Best} & \textbf{17.91} & \textbf{9.30} & \textbf{13.54} & \textbf{9.13} \\
\bottomrule
\end{tabular}
\vspace{0.5em}
\end{table}

%% file: table/Translate-EN.tex
\begin{table*}[t]
\centering
\caption{English translation benchmark results. All systems are evaluated on translation into English. sacreBLEU, chrF++, COMET, and BLEURT are reported, where higher scores indicate better performance.}
\label{tab:english_translation_results}
\renewcommand\tabcolsep{2pt}
\resizebox{\linewidth}{!}{
\begin{tabular}{llcccccccccccc}
\toprule[1pt]

Model & Metric
& ARE & DZA & EGY & IDN & IRQ & MAR & MYS & PHL & SAU & THA & VNM & AVG \\
\midrule

\multirow{4}{*}{azure\_trans}
& sacreBLEU & 13.80 & 16.04 & 20.07 & 19.27 & 21.43 & 13.95 & 13.80 & 17.29 & 26.27 & 16.42 & 25.73 & 18.55 \\
& chrF++ & 33.08 & 37.47 & 41.17 & 41.50 & 44.07 & 35.33 & 33.95 & 38.84 & 49.55 & 37.30 & 48.24 & 40.05 \\
& COMET & 0.60 & 0.62 & 0.64 & 0.69 & 0.67 & 0.58 & 0.64 & 0.67 & 0.71 & 0.69 & 0.72 & 0.66 \\
& BLEURT & 0.49 & 0.51 & 0.52 & 0.57 & 0.54 & 0.47 & 0.51 & 0.56 & 0.59 & 0.53 & 0.60 & 0.54 \\

\midrule

\multirow{4}{*}{gemini-3-flash-preview}
& sacreBLEU & 21.48 & 24.90 & 28.89 & 29.86 & 25.37 & 22.67 & 24.35 & 32.40 & 34.83 & 25.49 & 34.81 & 27.73 \\
& chrF++ & 41.12 & \textbf{47.50} & \textbf{50.99} & 52.19 & 49.99 & \textbf{45.35} & \textbf{44.89} & \textbf{55.81} & 57.68 & 46.42 & 56.08 & 49.82 \\
& COMET & \textbf{0.68} & \textbf{0.70} & \textbf{0.72} & 0.75 & 0.72 & \textbf{0.67} & 0.69 & \textbf{0.76} & 0.77 & 0.74 & 0.78 & 0.73 \\
& BLEURT & \textbf{0.57} & \textbf{0.60} & \textbf{0.61} & 0.64 & 0.61 & \textbf{0.57} & \textbf{0.57} & \textbf{0.66} & 0.67 & 0.59 & 0.67 & \textbf{0.61} \\

\midrule

\multirow{4}{*}{qwen35omniplus\_ast}
& sacreBLEU & \textbf{23.26} & \textbf{26.28} & \textbf{31.30} & \textbf{32.96} & \textbf{30.37} & \textbf{23.18} & \textbf{24.92} & \textbf{34.27} & \textbf{38.55} & \textbf{27.19} & \textbf{37.48} & \textbf{29.98} \\
& chrF++ & \textbf{41.54} & 46.51 & 50.33 & \textbf{53.71} & \textbf{51.06} & 43.51 & 43.81 & 54.08 & \textbf{59.07} & \textbf{47.22} & \textbf{57.73} & \textbf{49.87} \\
& COMET & 0.68 & 0.69 & 0.71 & \textbf{0.77} & \textbf{0.73} & 0.66 & \textbf{0.70} & 0.75 & \textbf{0.78} & \textbf{0.75} & \textbf{0.79} & \textbf{0.73} \\
& BLEURT & 0.56 & 0.57 & 0.59 & \textbf{0.65} & \textbf{0.61} & 0.54 & 0.56 & 0.63 & \textbf{0.67} & \textbf{0.60} & \textbf{0.68} & 0.61 \\

\midrule

\multirow{4}{*}{seamless\_m4t\_v2\_large}
& sacreBLEU & 8.26 & 14.77 & 15.97 & 17.87 & 16.60 & 13.97 & 13.18 & 16.87 & 21.72 & 11.31 & 16.70 & 15.20 \\
& chrF++ & 24.30 & 33.94 & 34.46 & 38.09 & 36.28 & 33.15 & 30.61 & 35.77 & 42.85 & 28.90 & 37.69 & 34.19 \\
& COMET & 0.55 & 0.62 & 0.60 & 0.66 & 0.63 & 0.60 & 0.60 & 0.64 & 0.68 & 0.62 & 0.68 & 0.63 \\
& BLEURT & 0.45 & 0.51 & 0.49 & 0.54 & 0.51 & 0.50 & 0.47 & 0.53 & 0.57 & 0.49 & 0.54 & 0.51 \\

\bottomrule
\end{tabular}
}
\vspace{-1em}
\end{table*}

%% file: table/Translate-CH.tex
\begin{table*}[t]
\centering
\caption{Chinese translation benchmark results. All systems are evaluated on translation into Chinese. sacreBLEU, chrF++, COMET, and BLEURT are reported, where higher scores indicate better performance.}
\label{tab:chinese_translation_results}
\renewcommand\tabcolsep{2pt}
\resizebox{\linewidth}{!}{
\begin{tabular}{llcccccccccccc}
\toprule[1pt]

Model & Metric
& ARE & DZA & EGY & IDN & IRQ & MAR & MYS & PHL & SAU & THA & VNM & AVG \\
\midrule

\multirow{4}{*}{azure\_trans}
& sacreBLEU & 14.46 & 17.20 & 21.47 & 21.99 & 23.57 & 15.28 & 17.45 & 19.16 & 26.75 & 17.67 & 26.96 & 20.18 \\
& chrF++ & 10.63 & 12.09 & 14.70 & 15.36 & 15.83 & 10.84 & 12.03 & 13.41 & 17.76 & 12.61 & 17.82 & 13.92 \\
& COMET & 0.60 & 0.63 & 0.64 & 0.70 & 0.67 & 0.58 & 0.67 & 0.68 & 0.70 & 0.69 & 0.75 & 0.67 \\
& BLEURT & 0.41 & 0.43 & 0.46 & 0.53 & 0.49 & 0.37 & 0.49 & 0.51 & 0.53 & 0.48 & 0.57 & 0.48 \\

\midrule

\multirow{4}{*}{gemini-3-flash-preview}
& sacreBLEU & 24.49 & \textbf{28.35} & \textbf{32.91} & 35.52 & 29.52 & \textbf{26.74} & \textbf{30.01} & \textbf{35.80} & 36.99 & \textbf{31.07} & 41.39 & \textbf{32.07} \\
& chrF++ & 16.51 & \textbf{18.71} & \textbf{21.85} & 23.54 & 20.47 & \textbf{17.55} & \textbf{19.61} & \textbf{23.69} & 24.13 & 20.18 & 26.81 & 21.19 \\
& COMET & \textbf{0.71} & \textbf{0.72} & \textbf{0.74} & 0.77 & 0.74 & \textbf{0.70} & 0.71 & \textbf{0.78} & 0.78 & 0.76 & 0.82 & \textbf{0.75} \\
& BLEURT & \textbf{0.53} & \textbf{0.56} & \textbf{0.58} & 0.61 & 0.58 & \textbf{0.53} & \textbf{0.54} & \textbf{0.63} & 0.63 & 0.57 & \textbf{0.67} & \textbf{0.59} \\

\midrule

\multirow{4}{*}{qwen35omniplus\_ast}
& sacreBLEU & \textbf{24.60} & 27.02 & 32.40 & \textbf{35.68} & \textbf{32.31} & 23.59 & 27.39 & 34.79 & \textbf{39.14} & 30.56 & \textbf{42.31} & 31.80 \\
& chrF++ & \textbf{16.59} & 18.43 & 21.78 & \textbf{24.65} & \textbf{21.64} & 16.33 & 18.60 & 23.26 & \textbf{25.82} & \textbf{20.84} & \textbf{28.19} & \textbf{21.47} \\
& COMET & 0.70 & 0.71 & 0.72 & \textbf{0.79} & \textbf{0.75} & 0.68 & \textbf{0.72} & 0.76 & \textbf{0.79} & \textbf{0.77} & \textbf{0.83} & 0.75 \\
& BLEURT & 0.52 & 0.54 & 0.55 & \textbf{0.63} & \textbf{0.58} & 0.50 & 0.53 & 0.60 & \textbf{0.64} & \textbf{0.58} & 0.67 & 0.57 \\

\midrule

\multirow{4}{*}{seamless\_m4t\_v2\_large}
& sacreBLEU & 2.53 & 8.90 & 7.59 & 11.47 & 10.30 & 8.36 & 8.91 & 11.12 & 11.62 & 8.18 & 12.68 & 9.24 \\
& chrF++ & 3.43 & 6.80 & 5.87 & 8.67 & 7.47 & 6.58 & 6.70 & 8.62 & 8.61 & 7.16 & 9.45 & 7.21 \\
& COMET & 0.48 & 0.58 & 0.53 & 0.60 & 0.57 & 0.56 & 0.57 & 0.57 & 0.59 & 0.57 & 0.62 & 0.57 \\
& BLEURT & 0.25 & 0.34 & 0.29 & 0.37 & 0.33 & 0.32 & 0.32 & 0.34 & 0.37 & 0.33 & 0.40 & 0.33 \\

\bottomrule
\end{tabular}
}
\vspace{-1em}
\end{table*}

%% file: text/bench.tex
\section{Experimental Results}

\subsection{Evaluated Systems}
We evaluate a comprehensive set of state-of-the-art ASR systems spanning both commercial APIs and open-source models.
Specifically, the commercial APIs include Microsoft Azure Speech\footnote{\url{https://portal.azure.com/\#view/Microsoft_Azure_ProjectOxford/CognitiveServicesHub/\~/SpeechServices}}, Google Chirp3\footnote{\url{https://docs.cloud.google.com/speech-to-text}}, OpenAI GPT-4o Transcribe\footnote{\url{https://developers.openai.com/api/docs/models/gpt-4o-transcribe}}, Gemini 3.0 Flash\footnote{\url{https://docs.cloud.google.com/gemini-enterprise-agent-platform/models/gemini/3-flash}}, ElevenLabs\_Scribe\_v2\footnote{\url{https://elevenlabs.io/}}, Qwen3-ASR-Flash~\footnote{\url{https://qwen.ai/blog?id=41e4c0f6175f9b004a03a07e42343eaaf48329e7&from=research.latest-advancements-list}}, Qwen3.5-Omni-plus\footnote{\url{https://qwen.ai/blog?id=qwen3.5-omni}}, Seed-ASR-1 (BIGASR\_V400)\footnote{\url{https://docs.byteplus.com/en/docs/byteplusvoice/asraudiofile}}, and Seed-ASR 2.0\footnote{\url{https://docs.byteplus.com/zh-CN/docs/byteplusvoice/speechtotextv2}};
the open-source models include Qwen3-ASR (1.7B)\footnote{\url{https://huggingface.co/Qwen/Qwen3-ASR-1.7B}}, Whisper-large-v3\footnote{\url{https://huggingface.co/openai/whisper-large-v3}}, NVIDIA NeMo Canary\footnote{\url{https://huggingface.co/nvidia/canary-1b}}, Meta OmniASR-LLM-3B\footnote{\url{https://huggingface.co/facebook/omniASR-LLM-3B}}, FunASR-MLT-nano~\footnote{\url{https://huggingface.co/FunAudioLLM/Fun-ASR-MLT-Nano-2512}}, and Dolphin (base\footnote{\url{https://modelscope.cn/models/DataoceanAI/dolphin-base}}/small\footnote{\url{https://modelscope.cn/models/DataoceanAI/dolphin-small}}).
For the speech-to-text translation module, we additionally evaluate Azure Translate, Gemini 3.0 Flash\footnote{\url{https://docs.cloud.google.com/gemini-enterprise-agent-platform/models/gemini/3-flash}}, SeamlessM4T-v2-Large\footnote{\url{https://huggingface.co/facebook/seamless-m4t-v2-large}}, and Qwen3.5-Omni-plus.
These systems are evaluated using the OpenSTBench ~\cite{an2026openstbenchbeyondsemanticevaluation} toolkit with standard translation-quality metrics, including sacreBLEU~\cite{post-2018-sacreBLEU}, chrF++~\cite{popovic-2017-chrf++}, COMET~\cite{rei2020cometneuralframeworkmt}, and BLEURT~\cite{sellam-etal-2020-bleurt}.
For each subsequent module, the same set of systems is evaluated wherever the language or domain is supported by the system; cells marked ``--'' in the result tables indicate that the corresponding system does not officially support that language or returned no usable output.

\subsection{Low Resource Language}
We construct a multilingual ASR benchmark targeting low-resource languages, with a particular focus on underrepresented regions across the Middle East, Southeast Asia, and East Asia. 
Although these regions collectively encompass over one billion speakers, they remain severely underrepresented in existing foundational ASR evaluations. 
Specifically, our benchmark covers seven Arabic-speaking regions: Iraq, Algeria, the United Arab Emirates, Egypt, Morocco, Saudi Arabia, and Syria; five Southeast Asian languages: Indonesian, Malay, Filipino (Tagalog), Vietnamese, and Thai; and two East Asian languages: Japanese and Korean. For Tagalog, we further provide two distinct subsets: one containing pure Tagalog and another capturing natural Tagalog-English code-switching. 

All evaluation data are collected from YouTube and manually curated to ensure high quality. Compared to traditional benchmarks that rely on read speech, our dataset better reflects real-world scenarios, featuring multi-speaker conversations and complex acoustic environments. Moreover, all audios are sourced from within the past year, which minimizes the risk of overlapping with existing training data and improves the fairness of the benchmark. 

Table~\ref{tab:low_resource_main} presents the performance of representative state-of-the-art ASR systems on our benchmark. Overall, these results show that current ASR systems still struggle substantially on newly collected in-the-wild low-resource and regional speech, even when they perform well on standard benchmarks such as Common Voice and FLEURS.

\subsection{Accented English}
To evaluate ASR robustness against pronunciation variability, we construct a test set comprising 10 hours of speech for each of six diverse English accents: Chinese, Indian, Japanese, Filipino, Scottish, and Singaporean. This selection strategically encompasses widely spoken second-language (L2) varieties alongside challenging localized and native variants. Consistent with our temporal hold-out strategy, these recordings are newly collected to mitigate data contamination. As shown in Table~\ref{tab:english_accents}, despite recent advancements in foundation models, substantial performance gaps remain between standard and accented English, underscoring the persistent challenge of geographical robustness.

\subsection{Chinese Dialects}

Our benchmark includes 10 hours of speech for each of the six major Chinese dialects: Xiang, Jin, Min, Yue, Gan and Wu. All dialects are provided with phonetic transcriptions, while Min is instead provided with Mandarin translations due to the extreme complexity of its character-to-pronunciation mapping. It is worth noting that the Yue and Wu dialect evaluation sets are sourced from WenetSpeech-YUE~\cite{wenetspeech-yue} and WenetSpeech-WU~\cite{wenetspeech-wu} datasets, while the remaining dialects are newly curated. The results demonstrate that Chinese dialect recognition remains highly challenging, especially for dialects with large phonological and writing-system divergence from Mandarin.

\subsection{Vertical Domains}

Standard open-domain evaluation often obscures models' vulnerabilities to dense, specialized vocabulary. To address this, we compile evaluation sets across twelve vertical domains: Agriculture, AI Technology, Arts, Biotechnology, E-commerce, Engineering, Entertainment, Finance, Humanities, Law, Medicine, and Military. For each domain, we manually curate 10 hours of parallel Chinese and English data sourced from major video platforms like YouTube, strictly adhering to the recent-year temporal constraint. To rigorously assess the recognition of domain-specific terminology, we extract technical keywords using the Qwen3~\cite{yang2025qwen3} large language model, followed by manual human verification.
Consequently, we report both the standard WER and the Biased Word Error Rate~(B-WER) to provide a granular evaluation of how effectively current systems recognize critical, long-tail entities.
For each utterance, we first use Qwen3.6-Max Preview\footnote{\url{https://qwen.ai/blog?id=qwen3.6-max-preview}} to annotate entity words in the reference transcription. 
The entity annotations are applied only to the reference transcription, while the ASR hypothesis is not entity-annotated. 
B-WER is then computed by applying the standard WER calculation only to reference tokens that belong to the annotated entity list:
\begin{equation}
\mathrm{B\text{-}WER}
=
\frac{S_b + D_b + I_b}{N_b}
\times 100\%
\end{equation}
where $N_b$ denotes the number of reference tokens belonging to the annotated entities, and $S_b$, $D_b$, and $I_b$ denote the numbers of substitution, deletion, and insertion errors associated with these entity tokens, respectively. 
Entity tokens that do not appear in the reference transcription are excluded from the B-WER denominator.
Due to space limitations, the standard CER and WER tables together with their detailed discussion are deferred to Appendix~\ref{sec:appendix_cer_wer}, and the B-CER results for Chinese and B-WER results for English are shown in Table~\ref{tab:vertical_domain_chinese_bcer} and Table~\ref{tab:vertical_domain_english_bwer}.  These results suggest that current ASR systems remain weak in recognizing specialized terms.

\subsection{Elderly and Children’s Speech}

Children’s and older adults’ speech differs substantially from standard adult speech in both acoustic characteristics and pronunciation patterns. Child speech is typically characterized by a higher fundamental frequency, developing vocal-tract characteristics, and less stable pronunciation, whereas older-adult speech may exhibit reduced volume, slower articulation, and age-related voice variation. To evaluate whether current ASR systems can handle such demographic differences, we construct dedicated child and older-adult speech evaluation sets for both English and Chinese, with 10 hours of speech for each age group in each language. The performance gap on child and older-adult speech suggests that demographic acoustic variation remains under-addressed by current ASR systems.

\subsection{Speech Translation}
GigaSpeechBench further supports speech-to-text translation evaluation by providing human-translated English and Chinese references for 11 underrepresented source languages. We evaluate Azure Translate, Gemini-3-Flash-Preview, SeamlessM4T-v2-Large, and Qwen3.5-Omni-plus~\cite{qwen35omni}, using sacreBLEU~\cite{post-2018-sacreBLEU}, chrF++~\cite{popovic-2017-chrf++}, COMET~\cite{rei2020cometneuralframeworkmt}, and BLEURT~\cite{sellam-etal-2020-bleurt}.
As shown in Table~\ref{tab:english_translation_results} and Table~\ref{tab:chinese_translation_results}, results show that translating in-the-wild low-resource and regional speech remains difficult.

%% file: text/conclusion.tex
\section{Conclusion}

In this work, we present GigaSpeechBench, a unified multilingual ASR benchmark designed to place long-tail evaluation axes on a common testbed. To construct this benchmark, we curated over 690 hours of newly collected, manually transcribed in-the-wild speech covering 12 underrepresented languages from the Middle East and Southeast Asia, six Chinese dialects, six English accents, and 12 hotword-rich vertical domains in both Chinese and English. 
Across this testbed, we conducted a large-scale comparison of representative commercial APIs, closed-source ASR-specialized models, and open-source foundation systems. Our evaluation reveals consistent performance gaps between standard academic datasets and more realistic conditions. We demonstrate that systems nearing saturation on existing standard benchmarks degrade substantially on in-the-wild low-resource and dialectal speech. Furthermore, accent robustness varies sharply across English varieties, and aggregate WER often masks sizeable degradation on entity-rich utterances in vertical domains.

We release GigaSpeechBench alongside its annotation protocols, hotword lists, and evaluation scripts. We hope this benchmark serves as a reproducible diagnostic tool for tracking progress on these underrepresented, yet practically critical, dimensions of robust speech recognition.

%% file: text/limitations.tex
\section*{Limitations}
Despite our effort to cover multilingual, multi-dialectal, and multi-domain speech in real-world scenarios, several limitations remain.
First, text normalization for low-resource languages still leaves room for further refinement by native-speaking linguistic experts.
Second, some Chinese dialects lack unified standard writing systems, so dialectal fragments are often approximately transliterated according to pronunciation.
As these transliterated fragments may admit multiple reasonable surface forms, different ASR systems may produce inconsistent character-level outputs for the same dialectal speech segment, making CER insufficient to fully assess whether a model captures the intended pronunciation, meaning, or dialectal expression.
Future work should therefore explore evaluation criteria better suited to low-resource languages and dialectal scenarios than WER/CER.

%% file: text/ethicsstatement.tex
\section*{Ethics Statement}
All collected audio is sourced from materials released under a Creative Commons license.
Samples containing personally identifiable information (PII) have been manually removed during the annotation process.
All annotators are compensated fairly through a professional data annotation company.
We are committed to ongoing maintenance of the dataset to address any potential risks in the future.

%% file: text/appendix.tex
\section{Full Author Affiliations} 
\begin{enumerate}[itemsep=0pt, topsep=2pt, leftmargin=*]
   \item X-LANCE Lab, MoE Key Lab of Artificial Intelligence, Jiangsu Key Lab of Language Computing, Shanghai Jiao Tong University
   \item Shanghai Innovation Institute
   \item Alibaba Group
   \item Tianjin University
   \item Tsinghua University
   \item Audio, Speech and Language Processing Group, School of Computer Science, Northwestern Polytechnical University
   \item Nanyang Technological University
   \item Institute of Automation, Chinese Academy of Sciences
   \item University of Chinese Academy of Sciences
   \item University of Illinois Urbana-Champaign, Urbana
   \item The Chinese University of Hong Kong, Shenzhen
   \item Fudan University, Shanghai, China.
   \item State Key Laboratory of Complex \& Critical Software Environment
   \item Seasalt.ai, Seattle
   \item WeNet Community
   \item SpeechColab
\end{enumerate}

\section{Segment Duration and Text Length Statistics}
\label{sec:appendix_duration}

\begin{figure}[t]
\centering
\includegraphics[width=0.8\linewidth]{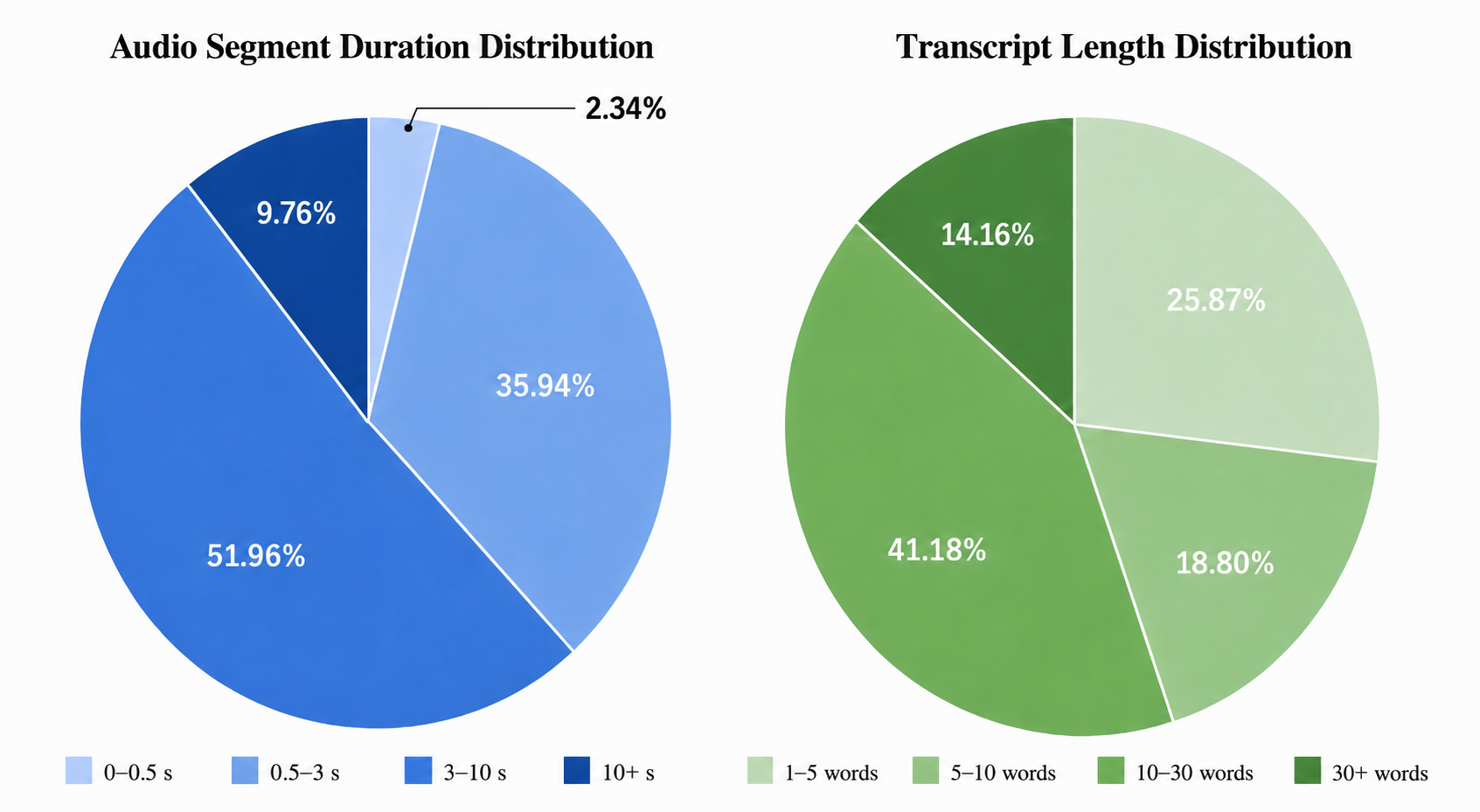}
\caption{
Distribution of audio segment duration and transcript length.
}
\vspace{-1em}
\label{fig:distribution}
\end{figure}

Figure~\ref{fig:distribution} shows the distributions of audio segment duration and reference text length after VAD and manual transcription. Overall, the data are concentrated in short-to-medium utterances, with the vast majority of audio segments falling between 0.5 and 10 seconds. This pattern is consistent with spontaneous conversational speech in real-world videos, where speech typically appears as natural turns or phrase-level segments rather than long, continuous read passages. Meanwhile, segments longer than 10 seconds are still retained in a non-negligible proportion, allowing the benchmark to cover not only short conversational turns but also longer utterances with richer contextual continuity. Extremely short segments below 0.5 seconds account for only 2.34\% of the data. Therefore, excluding them from the final metric computation helps avoid unstable error estimates caused by overly short references, while having little impact on the overall data distribution.

The reference text length distribution further supports this observation: most references fall within the 10--30 word range, indicating that the dataset is primarily composed of natural utterance-level speech rather than overly long discourse-level recordings. The presence of both shorter and longer references reflects the natural variation in expression length in real-world speech. In addition, we still provide the complete audio recordings to support potential future evaluation on long-form speech recognition.

\section{CER and WER results of Vertical Domain}
\label{sec:appendix_cer_wer}

\input{table/Chinese-CER}
\input{table/English-WER}
Table 12 and Table 13 provide supplementary standard CER and WER results on the twelve vertical domains. Overall, the standard error rates are generally lower than the entity-level B-CER and B-WER reported in the main text, indicating that current systems can achieve relatively stable performance in general transcription. However, such aggregate metrics mainly reflect average transcription quality and can be dominated by non-terminological words, making them insufficient to fully reveal recognition errors on domain-specific terminology and long-tail entities. In contrast, the entity-level evaluation in the main text offers a more fine-grained diagnostic perspective and better reflects the robustness of ASR systems in real-world vertical-domain scenarios.

%% file: table/Chinese-CER.tex
\begin{table*}[!t]
\centering
\caption{Chinese vertical-domain benchmark results, reported in CER (\%).}
\renewcommand\tabcolsep{2pt}
\resizebox{\linewidth}{!}{
\begin{tabular}{lcccccccccccc}
\toprule[1pt]

\textbf{Model}
& AGR-CH
& AIT-CH
& ART-CH
& BIO-CH
& ECM-CH
& ENG-CH
& ENT-CH
& FIN-CH
& HUM-CH
& LAW-CH
& MED-CH
& MIL-CH \\

\midrule

Azure
& 7.31 & 5.55 & 5.30 & 6.09 & 11.91 & 7.34 & 6.05 & 2.75 & 5.88 & 6.38 & 3.80 & 2.66 \\

Chirp 3
& 9.63 & 8.98 & 7.40 & 6.45 & 15.63 & 8.58 & 9.26 & 4.55 & 5.84 & 26.44 & 4.83 & 5.02 \\

ElevenLabs Scribe v2
& 6.61 & 5.22 & 4.03 & 4.84 & 12.25 & 5.94 & 5.94 & 2.52 & 3.51 & 6.38 & 3.24 & 2.36 \\

Meta OmniASR 3B
& 14.07 & 14.98 & 10.58 & 12.73 & 18.85 & 14.42 & 11.61 & 7.49 & 9.45 & 10.87 & 9.63 & 7.52 \\

Qwen3-ASR-Flash
& 4.84 & 5.37 & 9.21 & 5.25 & 11.36 & 5.13 & 4.92 & 2.88 & 2.32 & 10.33 & 6.88 & 5.92 \\

Qwen3-ASR-1.7B
& 4.65 & 4.42 & 2.95 & 2.81 & 10.01 & 4.18 & 4.76 & 1.82 & 2.45 & 5.20 & 2.27 & 1.91 \\

NVIDIA NeMo
& 29.95 & 36.11 & 23.80 & 26.53 & 38.86 & 29.07 & 31.71 & 20.30 & 22.06 & 22.75 & 21.75 & 20.47 \\

GPT-4o Transcribe
& 15.50 & 19.58 & 15.11 & 11.58 & 29.30 & 29.63 & 12.61 & 13.32 & 7.08 & 13.58 & 8.73 & 7.41 \\

Gemini 3.0 Flash
& 10.92 & 9.92 & 7.85 & 6.44 & 18.42 & 9.19 & 11.05 & 5.57 & 5.39 & 10.79 & 4.79 & 5.20 \\

Whisper Large v3
& 11.57 & 9.90 & 10.29 & 10.05 & 17.17 & 10.97 & 9.63 & 5.89 & 8.12 & 10.49 & 7.79 & 6.04 \\

Dolphin Small
& 9.64 & 9.54 & 6.29 & 8.01 & 12.46 & 7.85 & 8.74 & 3.53 & 6.86 & 6.97 & 5.64 & 5.11 \\

Dolphin Base
& 12.20 & 11.42 & 8.29 & 11.04 & 14.86 & 9.99 & 11.71 & 5.17 & 10.13 & 8.76 & 7.96 & 7.04 \\

FunASR-MLT-Nano
& 5.83 & 5.09 & 3.60 & 4.11 & 10.49 & 4.92 & 5.23 & 2.13 & 3.80 & 5.68 & 2.71 & 2.41 \\

FunASR-Realtime
& \textbf{3.15} & \textbf{3.33} & \textbf{2.60} & \textbf{2.07} & \textbf{9.10} & \textbf{2.65} & \textbf{3.41} & \textbf{1.44} & \textbf{1.68} & \textbf{4.97} & \textbf{1.56} & \textbf{1.49} \\

Qwen3.5-Omni-Plus
& \underline{3.54} & \underline{3.74} & \underline{2.65} & \underline{2.28} & \underline{9.35} & \underline{3.49} & \underline{3.59} & \underline{1.62} & \underline{1.71} & \underline{5.06} & \underline{1.70} & \underline{1.64} \\

BigASR
& 4.02 & 4.35 & 2.91 & 2.92 & 9.58 & 4.15 & 4.10 & 2.20 & 2.03 & 5.74 & 2.11 & 2.00 \\

SeedASR
& 4.02 & 4.33 & 2.91 & 2.89 & 9.62 & 4.18 & 4.05 & 2.17 & 2.03 & 5.73 & 2.09 & 2.02 \\

\midrule

\textbf{Best}
& \textbf{3.15} & \textbf{3.33} & \textbf{2.60} & \textbf{2.07} & \textbf{9.10} & \textbf{2.65} & \textbf{3.41} & \textbf{1.44} & \textbf{1.68} & \textbf{4.97} & \textbf{1.56} & \textbf{1.49} \\

\bottomrule[1pt]
\end{tabular}
}
\label{tab:vertical_domain_chinese}
\end{table*}

%% file: table/English-WER.tex
\begin{table*}[!t]
\centering
\caption{English vertical-domain benchmark results, reported in WER (\%).}
\renewcommand\tabcolsep{2pt}
\resizebox{\textwidth}{!}{
\begin{tabular}{lcccccccccccc}
\toprule[1pt]

\textbf{Model}
& AGR-EN
& AIT-EN
& ART-EN
& BIO-EN
& ECM-EN
& ENG-EN
& ENT-EN
& FIN-EN
& HUM-EN
& LAW-EN
& MED-EN
& MIL-EN \\
\midrule

Azure
& 7.00 & 10.58 & 6.37 & 6.94 & 10.94 & 7.40 & 9.84 & 8.18 & 7.34 & 10.79 & 6.06 & 6.48 \\

Chirp 3
& 6.92 & 10.46 & \underline{5.58} & 5.87 & 10.05 & 6.77 & 9.29 & 7.93 & 7.22 & 10.18 & 5.51 & 6.22 \\

ElevenLabs Scribe v2
& 13.61 & 15.13 & 10.80 & 9.89 & 15.37 & 10.27 & 18.44 & 13.76 & 12.21 & 16.82 & 9.87 & 11.43 \\

Meta OmniASR 3B
& 34.76 & 36.49 & 40.12 & 16.73 & 24.02 & 23.73 & 57.73 & 32.71 & 17.89 & 35.39 & 21.05 & 10.89 \\

Qwen3-ASR-Flash
& 6.63 & 11.26 & 11.86 & 6.78 & \textbf{8.78} & 5.63 & 9.21 & \underline{7.20} & 9.14 & 13.20 & 7.09 & 6.11 \\

Qwen3-ASR-1.7B
& \textbf{5.44} & \textbf{8.85} & \textbf{5.18} & 5.66 & 8.93 & \underline{5.29} & \textbf{8.09} & \textbf{6.24} & \underline{6.77} & \textbf{9.13} & \textbf{5.02} & \textbf{5.04} \\

NVIDIA NeMo
& 10.59 & 14.46 & 8.19 & 9.33 & 13.10 & 8.35 & 17.02 & 8.93 & 9.32 & 15.86 & 6.63 & 5.63 \\

GPT-4o Transcribe
& 21.86 & 34.51 & 18.87 & 9.77 & 35.10 & 24.20 & 30.51 & 20.53 & 17.53 & 21.96 & 30.62 & 14.84 \\

Gemini 3.0 Flash
& 8.36 & 12.09 & 6.66 & 6.38 & 11.25 & 6.49 & 12.54 & 8.65 & 7.56 & 13.26 & 5.82 & 6.22 \\

Whisper Large v3
& 9.43 & 13.56 & 7.29 & 6.29 & 10.52 & 6.67 & 11.61 & 9.37 & 7.66 & 11.25 & 5.69 & 6.08 \\

FunASR-MLT-Nano
& 8.16 & 11.87 & 7.31 & 7.09 & 9.87 & 6.23 & 11.50 & 8.40 & 7.87 & 11.80 & 5.91 & 5.99 \\

FunASR-Realtime
& \underline{6.39} & \underline{9.11} & \underline{5.58} & \underline{5.63} & \underline{8.86} & \textbf{4.88} & \underline{9.16} & 7.36 & \textbf{6.37} & 9.67 & 5.19 & \underline{5.26} \\

Qwen3.5-Omni-Plus
& 7.10 & 10.58 & 6.13 & \textbf{5.49} & 8.99 & 5.32 & 9.47 & 7.33 & 6.94 & \underline{9.20} & \underline{5.10} & 5.53 \\

BigASR
& 7.25 & 10.81 & 6.50 & 6.46 & 9.80 & 6.03 & 10.65 & 7.55 & 7.41 & 11.34 & 5.91 & 6.20 \\

SeedASR
& 9.14 & 11.61 & 6.54 & 6.50 & 9.79 & 5.99 & 10.63 & 7.54 & 7.47 & 11.39 & 5.90 & 6.20 \\

\midrule

\textbf{Best}
& \textbf{5.44} & \textbf{8.85} & \textbf{5.18} & \textbf{5.49} & \textbf{8.78} & \textbf{4.88} & \textbf{8.09} & \textbf{6.24} & \textbf{6.37} & \textbf{9.13} & \textbf{5.02} & \textbf{5.04} \\

\bottomrule[1pt]
\end{tabular}
}
\label{tab:vertical_domain_english}
\end{table*}